\documentclass[prl,twocolumn,superscriptaddress,floatfix,showpacs]{revtex4-1}
\usepackage{graphics,amssymb,amsmath,epsfig,color}

\newcommand{\be}{\begin{equation}} \newcommand{\ee}{\end{equation}}
\newcommand{\bea}{\begin{eqnarray}} \newcommand{\eea}{\end{eqnarray}}

\begin{document}

\title{Explosive Percolation is Continuous, but with Unusual Finite Size Behavior}
\author{Peter Grassberger} \affiliation{FZ J\"ulich, D-52425 J\"ulich, Germany} \affiliation{Complexity Science Group, University of Calgary, Calgary T2N 1N4, Canada}
\author{Claire Christensen} \affiliation{Complexity Science Group, University of Calgary, Calgary T2N 1N4, Canada}
\author{Golnoosh Bizhani} \affiliation{Complexity Science Group, University of Calgary, Calgary T2N 1N4, Canada}
\author{Seung-Woo Son} \affiliation{Complexity Science Group, University of Calgary, Calgary T2N 1N4, Canada}
\author{Maya Paczuski} \affiliation{Complexity Science Group, University of Calgary, Calgary T2N 1N4, Canada}
\date{\today}

\begin{abstract}
We study four Achlioptas type processes with ``explosive" percolation transitions.
All transitions are clearly continuous, but their finite size scaling 
functions are not entire holomorphic. The distributions of the 
order parameter, the relative size $s_{\rm max}/N$ of the largest cluster, are double-humped.
But -- in contrast to first order phase transitions -- the distance between the 
two peaks decreases with system size $N$ as $N^{-\eta}$ with $\eta > 0$.
We find different positive values of $\beta$ (defined via $\langle s_{\rm max}/N
\rangle \sim (p-p_c)^\beta$ for infinite systems) for each model, showing that they are all in different 
universality classes. In contrast, the exponent $\Theta$ (defined such that observables are 
homogeneous functions of $(p-p_c)N^\Theta$) is close to -- or even equal to -- 1/2 for all models.

\end{abstract}


\maketitle

Percolation is a pervasive concept in statistical physics and probability theory and has been 
studied {\it in extenso} in the past.
It came thus as a surprise to many, when Achlioptas {\it et al.}~\cite{Achli-2009} claimed that a 
seemingly mild modification of standard percolation models leads to a discontinuous phase transition
-- named ``explosive percolation" (EP) by them -- in contrast to the continuous phase transition seen in 
ordinary percolation. Following~\cite{Achli-2009} there appeared a flood of papers 
\cite{Friedman-2009,Ziff-2009,Ziff-2010,Radicchi-2009,Radicchi-2010,Souza-2010,Rozenf-2010,Manna-2009,Moreira-2010,Cho-2009,Cho-2009a,Cho-2010,Araujo-2010,Basu-2010,Tian-2010,Chen-2010,Nagler-2011,Hooy-2011,Costa-2010}
studying various aspects and generalizations of EP.
In all cases, with one exception~\cite{Costa-2010}, the authors agreed that the transition 
is discontinuous: the ``order parameter", defined as the fraction of vertices/sites in the 
largest cluster, makes a discrete jump at the percolation transition. In the present paper we 
join the dissenting minority and add further convincing evidence that the EP transition is {\it continuous}
in all models, but with unusual finite size behavior.

From the physical point of view, the model seems somewhat unnatural, since it involves non-local 
control (there is a `supervisor' who has to compare distant pairs of nodes to chose 
the actual bonds to be established~\footnote{In contrast to claims made by the authors, 
the model in~\cite{Souza-2010} is also non-local by any standard physics definition.}). 
Also, notwithstanding~\cite{Rozenf-2010}, no realistic applications have been proposed.
It is well known that the usual concept of universality classes in critical phenomena is invalidated 
by the presence of long range interactions. Thus it is not surprising that a percolation model 
with global control can show completely different behavior~\cite{Christen-2010}.

Usually, e.g. in thermal equilibrium systems, discontinuous phase 
transitions are identified with ``first order" transitions, while continuous transitions are
called ``second order". This notation is also often applied to percolative transitions. But EP 
lacks most attributes -- except possibly for the discontinuous order parameter 
jump -- considered essential for first order transitions. None of these other attributes (cooperativity, 
phase coexistence, and nucleation) is observed in Achlioptas type processes, although they are observed
in other percolation-type transitions~\cite{Janssen-2004}. 
Thus EP should never have been viewed as a first 
order transition, and it is gratifying that it is also not discontinuous. 

Apart from the behavior of the average value $\langle m\rangle$ of the order parameter $m$, phase 
transitions can also be characterized by the distribution $P_{p,N}(m)$ of $m$ in finite systems, 
where $p$ is the control parameter and $N$ measures the system size. For infinite $N$, $\langle m\rangle$ 
jumps at $p=p_c$ if the transition is discontinuous, while it varies continuously with a power
law singularity $\langle m\rangle \sim (p-p_c)^\beta$ for a continuous transition. 
The distribution $P_{p=p_c,N}(m)$ at criticality scales, for continuous transitions, as \cite{Bruce-1992}
\be
   P_{p=p_c,N}(m) \sim N^\eta f(m N^\eta),                            \label{P_m-critical}
\ee
where $\eta=\beta/(d\nu)$ for standard thermal second order phase transitions. The universal function $f(z)$
might be double-humped, as in the Ising model \cite{Bruce-1992}. But then, as $N\to\infty$, the dip 
between the humps usually does not deepen and the horizontal distance between them shrinks to zero
so that $P_{p=p_c,N}(m)$ becomes single-humped.

Equation~(1) is directly related to 
the finite size scaling (FSS) of $\langle m\rangle$ \cite{Binder-2010},
\be
   \langle m\rangle \sim (p-p_c)^\beta g[(p-p_c)N^\Theta],             \label{fss}
\ee
where the universal scaling function $g(z)$ is analytic at all finite $z$, reflecting the fact that the 
critical point was the only singularity of the partition function, before it was regularized by Eq.(\ref{fss}).
Notice that the usual FSS ansatz \cite{Binder-2010} involves the linear system size $L$ instead of $N$
with $\Theta=1/(d\nu)$, where $d$ is the dimension and $\nu$ is the correlation length exponent.

In typical first order transitions, in contrast, $P_{p=p_c,N}(m)$ is double-humped with a deepening valley
between the two peaks. The distance between the peaks tends to a positive constant which is equal to 
the jump in $\langle m\rangle$. The depth of the valley between the peaks reflects the fact that values
of $m$ between the peaks correspond to systems with two co-existing phases
and an interface between them that costs energy and is disfavored. As a consequence, systems
with first order transitions typically do not show FSS (unless the interface energy does not 
increase with system size \cite{Causo-2000}).

\begin{figure}
\begin{center}
\psfig{file=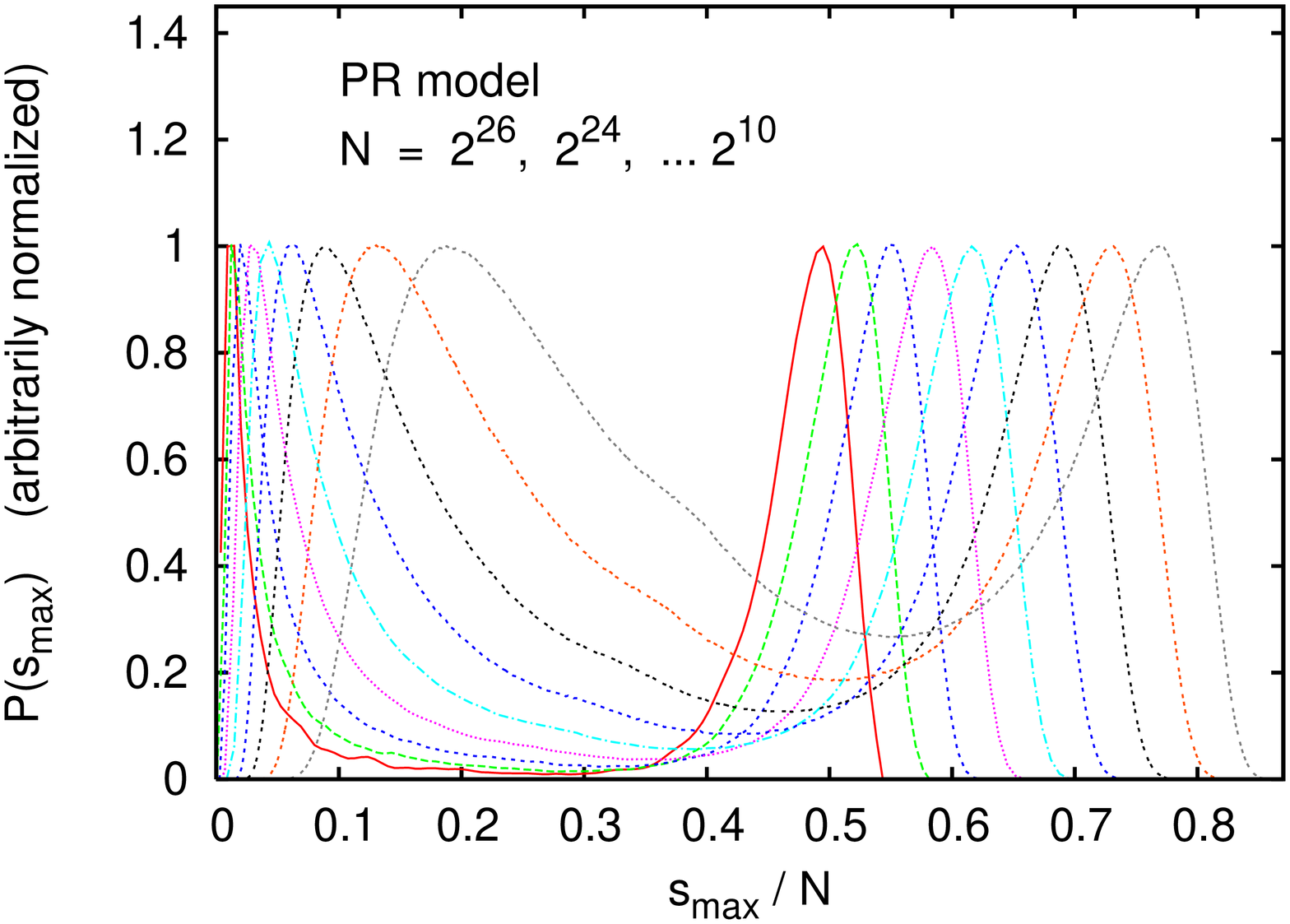,width=4.2cm, angle=270}
\psfig{file=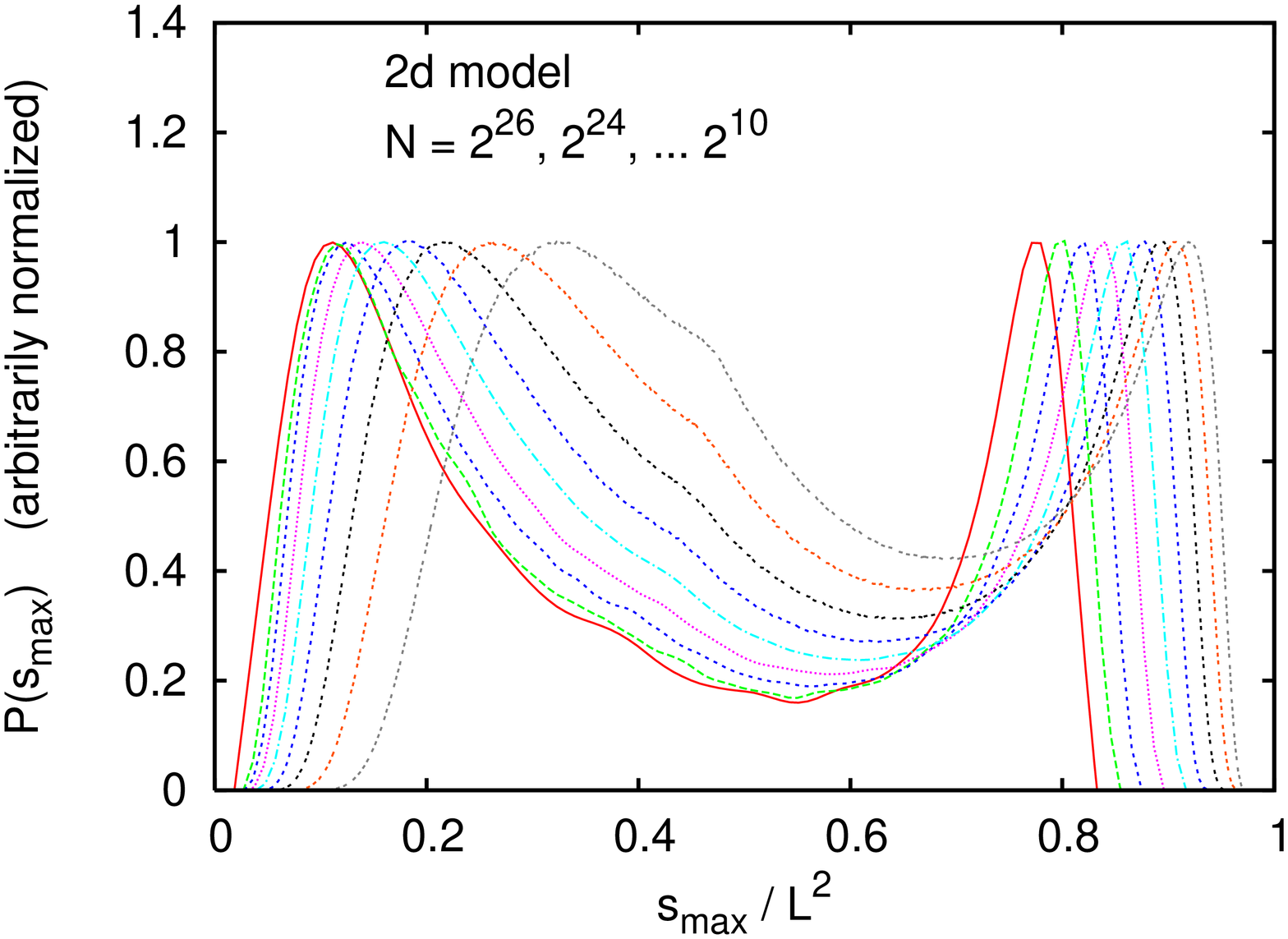,width=4.2cm, angle=270}
\psfig{file=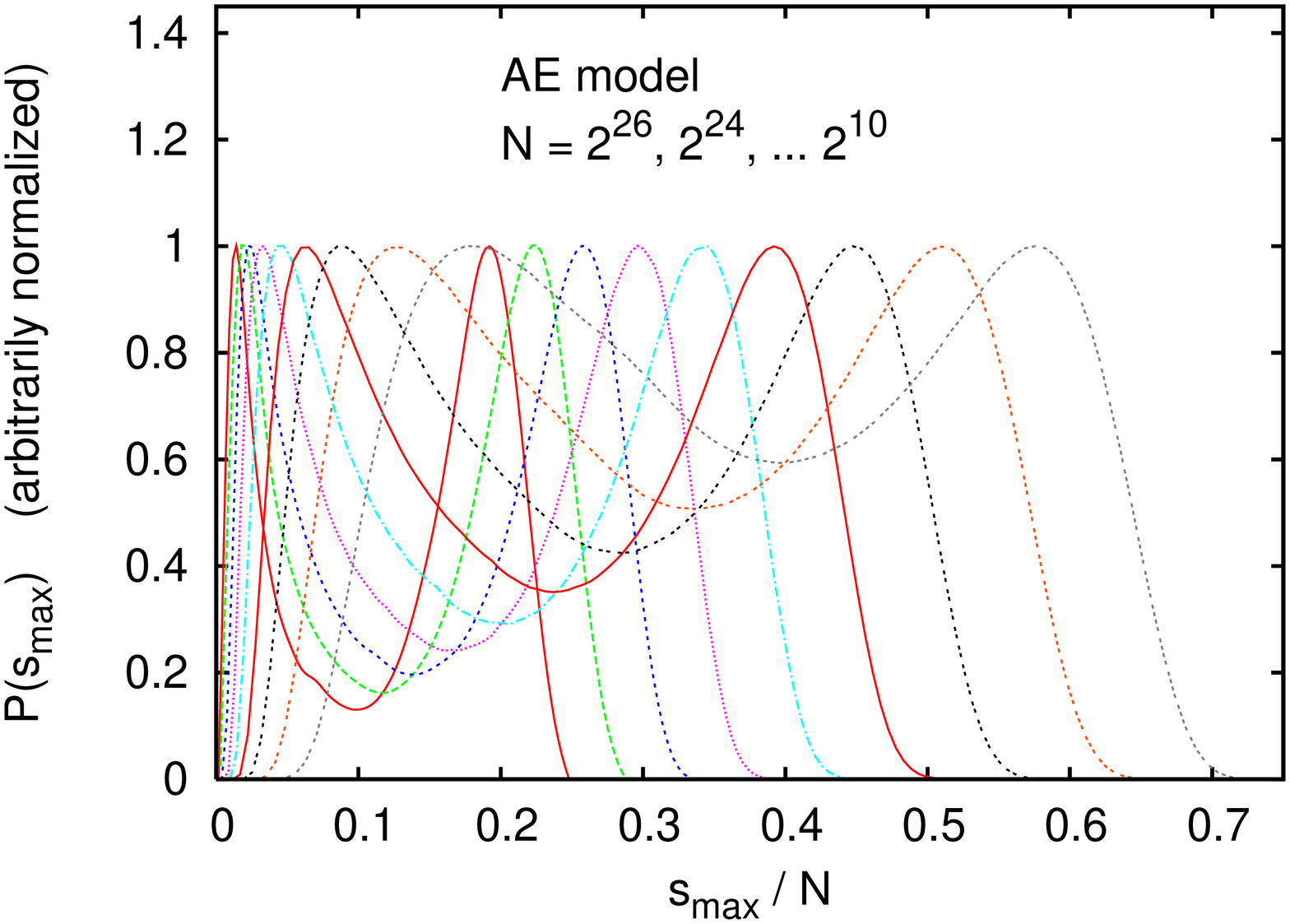,width=4.2cm, angle=270}
\psfig{file=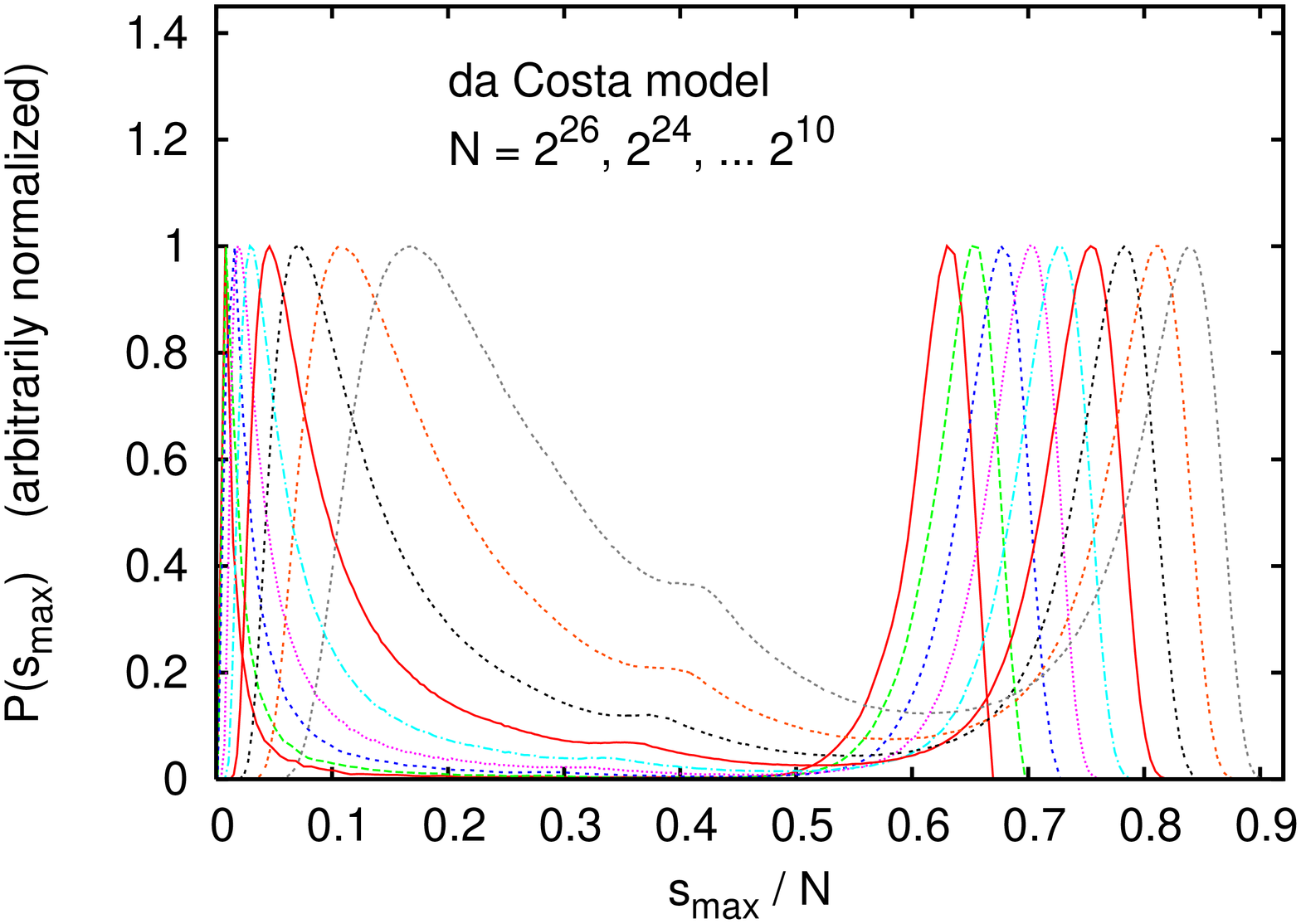,width=4.2cm, angle=270}
\caption{(Color online) Distributions of the order parameter $s_{\rm max}/N$ for four EP
models. They are shown at the effective critical point, defined such that both peaks have the same height. 
Normalization is such that their height is 1. For the largest systems, curves were approximated by cubic 
splines to make them smooth. }
\end{center}
\vglue -.3cm
\end{figure}

In percolation, usually the relative size of the largest cluster, $m\equiv s_{\rm max}/N$, is taken as
an order parameter. Here, $N$ is the number of nodes, and $s_{\rm max}/N\to 0$ for $p<p_c$ and $N\to\infty$.
In \cite{Radicchi-2010,Tian-2010} it was observed that $P_{p=p_c,N}(m)$ is strongly double-peaked in EP
transitions.  In \cite{Tian-2010} this 
was also backed by careful measurements of the depth of the valley between the peaks, which indeed lowered 
with increasing $N$. This was taken as a clear indication for 
the transition being first order and for phase coexistence. Notice that the latter
is not justified since $s_{\rm max}/N$ is, in contrast to the local order parameters in thermal systems,
a global quantity and cannot be used to characterize any part of a large system. Rather, the structure
of $P_{p=p_c,N}(m)$ in EP reflects the suddenness of the transition, combined with a scatter
of the precise $p$-values where individual systems acquire giant clusters. At $p$-values where both
peaks have the same height, it is much more likely to find either no giant cluster or a fully developed
one, than to find a half-grown giant cluster. Hence, the two peaks are more 
reminiscent of systems without self-averaging~\cite{Wise-1998} than of phase coexistence.

While the two peaks prove the suddenness of the transition that was claimed as a hallmark of EP, 
they do not yet prove that EP is discontinuous. For that, one must also show that the distance 
between the peaks does not vanish for $N\to\infty$. In order to check this, we have made 
extensive simulations of four models: The original product rule of \cite{Achli-2009}, denoted in the 
following as ``PR"; The product rule on 2-d square lattices \cite{Ziff-2009,Ziff-2010} with helical boundary 
conditions (``2d"); The `adjacent edge' rule \cite{Souza-2010} (``AE"); And the rule of \cite{Costa-2010} 
(``da Costa"). For more details on the simulations, see the supplementary material (SM).

\begin{table}
\begin{center}
\begin{tabular}{|c|l|l|l|l|} \hline
            &    PR      &     2d      &    AE      &  da Costa     \\ \hline
  $p_c$     & 0.888449(2)& 0.526562(3) & 0.797013(3) & 0.923207508  \\
  $\eta_+$  & 0.0402(15) &   0.018(2)  &  0.103(2)  &   0.0255(8)   \\
  $\eta_-$  &  0.270(7)  &   0.078(7)  &  0.228(5)  &   0.300(5)    \\
  $\beta$   &  0.0861(5) &   0.040(2)  &  0.214(2)  &   0.0557(5)   \\
 $\Theta_1$ &  0.47(2)   &   0.45(6)   &  0.48(1)   &   0.46(2)     \\ 
 $\Theta_2$ &  0.52(1)   &   0.47(3)   &  0.51(1)   &   0.53(1)     \\
 $\Theta_{\rm conj}$ & 1/2   &  --     &    1/2     &     1/2       \\
 $\eta_0$   &  0.0567(9) &   0.0612(8) &  0.1113(8) &   0.0356(8)   \\ \hline

\end{tabular}
\caption{Critical points and critical exponents for the four models. The $\Theta_i$ are 
different estimates of the exponent $\Theta$: $\Theta_1$ is obtained from the scaling relation
$\Theta = \eta_+/\beta$, $\Theta_2$ is obtained from a data collapse 
in the slightly supercritical region where $\langle m\rangle \approx m_+$,
and $\Theta_{\rm conj}$ is the conjectured exact value. For the da Costa model, $p_c$
is taken from \cite{Costa-2010}. For the other models it is obtained from plots analogous
to the inset in Fig.~4.} 
\end{center}
\end{table}

\begin{figure}
\begin{center}
\psfig{file=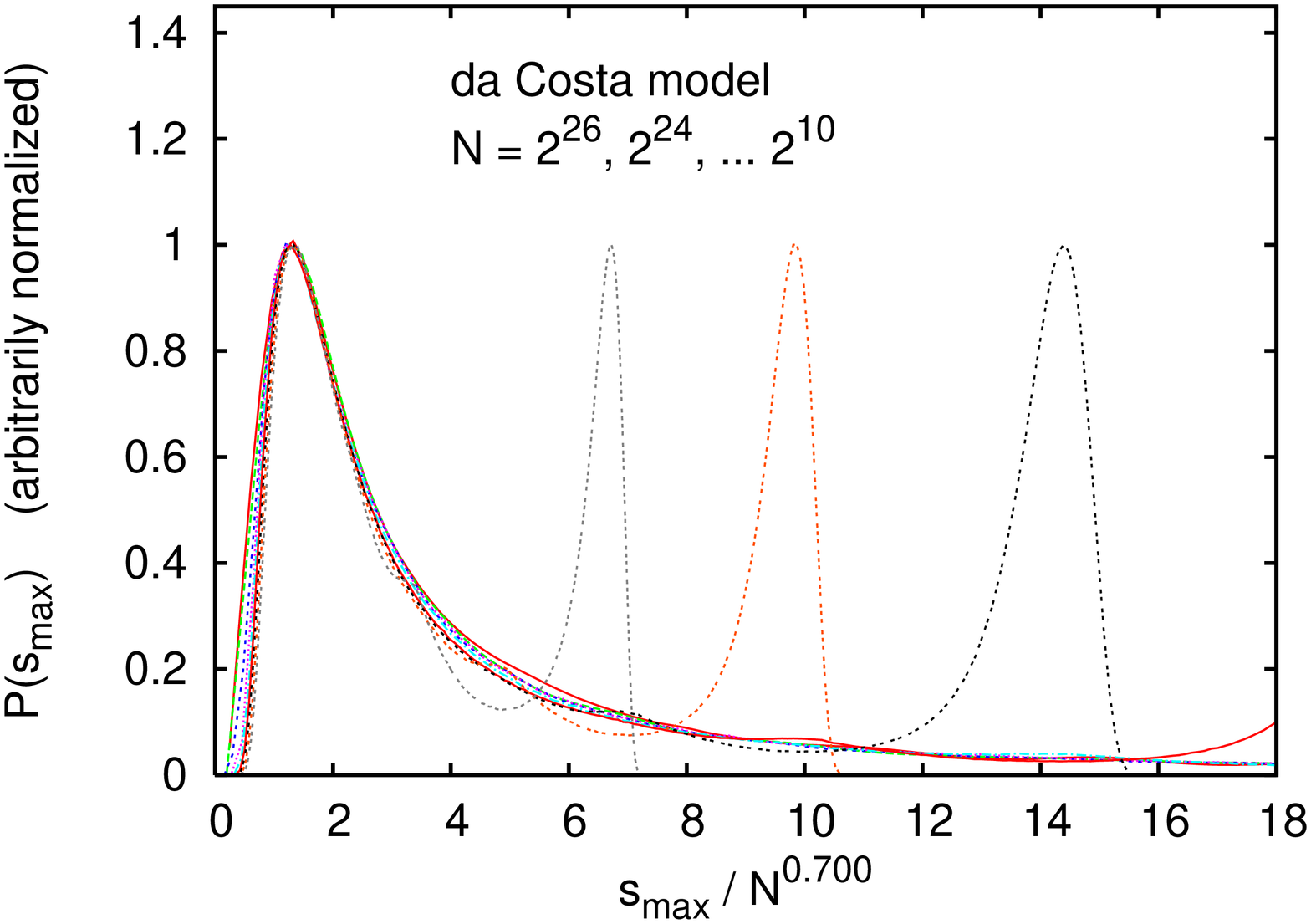,width=4.2cm, angle=270}
\psfig{file=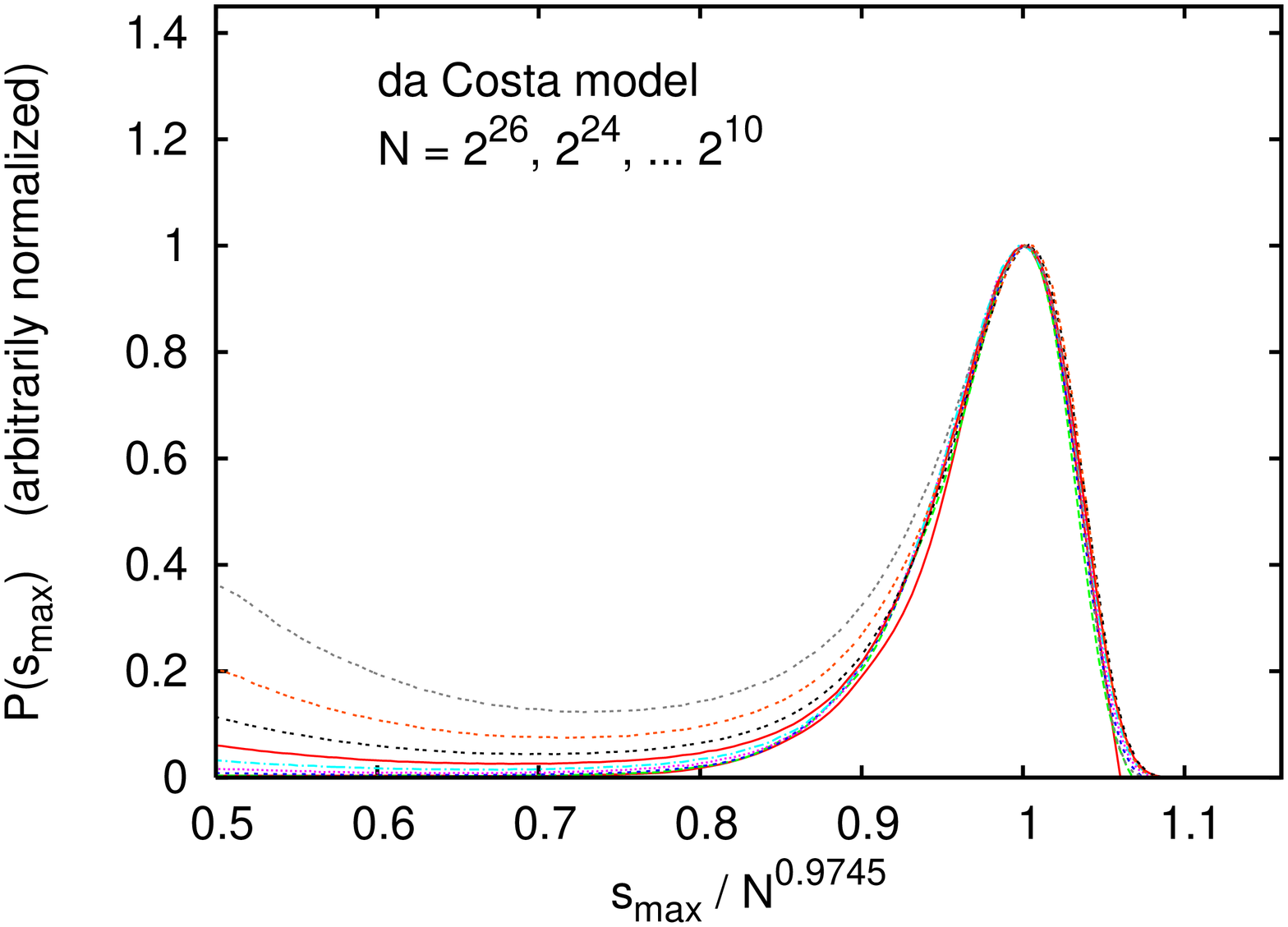,width=4.2cm, angle=270}
\caption{(Color online) Data collapses for the two peaks in the order parameter distribution for the 
da Costa model. Colors  and line styles are the same as in Fig.~1.}
\end{center}
\end{figure}

Distributions $P_{p,N}(m)$ for these models are shown in Fig.~1. In all cases $p$ was chosen such
that both peaks have equal height (set arbitrarily to 1). The extrapolations of these values for $N\to
\infty$ are given in Table 1. They agree within errors with the critical $p_c$ values quoted in 
the literature. We see that in each case the valley between the peaks deepens with increasing $N$ 
\cite{Tian-2010}, but at the same time both peaks shift to the left. 
Among the three off-lattice models, the AE model (the least non-local) shows the fastest
peak shifting and slowest valley deepening, while the opposite is true for the da Costa model.
In all cases this shift is compatible with power laws 
\vglue -9.mm
\be
   m_\pm \sim N^{-\eta_\pm},      \label{eta}
\ee
where $m_+\; (m_-)$ is the position of the right (left) peak at the critical point. In all cases 
$0 <\eta_+ < \eta_-$ (see Table 1), 
i.e. the right peak moves slower than the left one. Therefore the distance between the peaks increases 
for small $N$, but has finally to decrease $\sim N^{-\eta_+}$. Since this distance is asymptotically 
proportional to the maximum of the variance of $m$ \footnote{Here we use the observation 
\cite{Tian-2010} that the peaks mainly change their heights when $p$ is changed, and stay approximately
at the same position -- the slight shifts of their positions during this change might indeed be 
responsible for the differences between $\Theta_1$ and $\Theta_2$. 
}, 
we find that the variances first increase with $N$ (in agreement 
with \cite{Ziff-2010}), but ultimately must decrease. 

As shown in Fig.~2 for the da Costa model, 
not only the positions of the peaks scale, but also their widths. This indicates that the 
asymptotic scenario is two well separated peaks with $N-$independent shapes whose widths are 
proportional to their positions. If we switch from defining $p_c$ by equal peak heights to equal peak 
areas [28] and allow weak convergence for $N\to\infty$ (in contrast to the usual assumption 
of pointwise convergence; see SM) the full distributions at $p_c(N)$ then show asymptotic scaling
\vglue -7.mm
\be
 P_{p_c(N),N}(m) \sim N^{\eta_+} f(m N^{\eta_+})            \label{f-delta}
\ee 
with the scaling function $f(x)$ consisting of a finite width right hand peak and a $\delta-$peak at $x=0$.

\begin{figure}
\begin{center}
\psfig{file=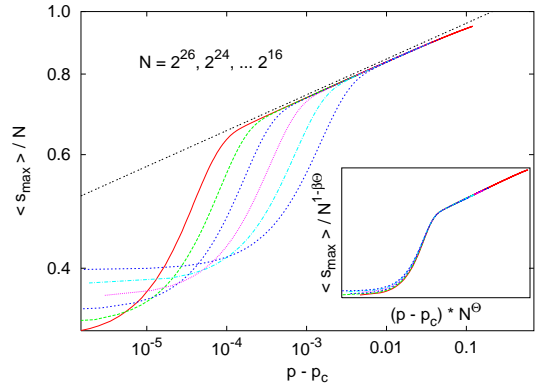,width=5.2cm, angle=270}
\caption{(Color online) Log-log plot of the average order parameter for the da Costa model {\it versus} $p-p_c$,
for six different values of $N$. One sees clearly a common part with slope $\beta$ (indicated also by the 
straight line), from which curves for different $N$ deviate later and later, as $N$ increases. The inset 
shows the collapse of these data as predicted by Eq.~(\ref{fss}). While $\Theta$ is fitted, both $\beta$ 
and $p_c$ are taken from \cite{Costa-2010}.}
\end{center}
\end{figure}

For $p$ strictly larger than $p_c(N)$, 
only the right hand peak dominates the average $\langle m\rangle$.
We then expect only small finite size scaling corrections to its asymptotic values, i.e. we 
expect the curves $\langle m\rangle_{p,N}$ for different $N$ to coincide for $p>p_c(N)$ on a common 
curve $\langle m\rangle_p$. Since the scenario in this regime is not much different from 
other critical phenomena this should be a power law $\langle m\rangle \sim (p-p_c)^\beta$ that 
holds in the range $m_+ < \langle m\rangle \ll 1$. Measured values of $\beta$ are given in Table~1. For 
the da Costa model the agreement with \cite{Costa-2010} is perfect.
Assuming Eq.~(\ref{fss}), it follows that $\Theta = \eta_+/\beta$. Values of $\Theta$ obtained from this, 
denoted as $\Theta_1$, are  slightly smaller than 1/2 for all models (see Table~1).

Deviations from this common power law are expected to set in when $\langle m\rangle$ decreases below 
$m_+$. The data for the da Costa model are shown in Figs.~3. For all $p>p_c$ (except for very small 
values of $z =(p-p_c)N^\Theta$), these deviations are fully described by the FSS ansatz in 
Eq.~(\ref{fss}). In Figs.~3 we chose $\Theta$ so that the collapse is best at $\langle m\rangle 
\approx m_+$, resulting in the value $\Theta_2$ quoted in Table~1. For the other models the data collapse 
is similarly good, except for the 2d model where it is worse (see SM). 
For all models, $\Theta_2$ is slightly larger than $\Theta_1$.

The fact that $f(z)$ in Eq.~(\ref{f-delta}) contains a $\delta$-peak at its leftmost extremity $z=0$ 
implies that $g(z)$ in Eq.(\ref{fss}) must vanish for all $z$ below some value $z_0 \leq 0 $, 
which in turn means that $g(z)$ must have a singularity at $z_0$.  Indeed, Fig.~4 shows that the values 
of $g(z)$ for $z<-1$ approach 0 very fast with increasing $N$, implying $-1 < z_0\leq 0$ (the latter is 
also true for the other models). We cannot exclude the possibility the curves in Fig.~4 approach 
a pure power law $az^\beta$ (dashed red line) in the limit $N\to\infty$.

The blow-up of the region around $z=0$ shown in the inset in Fig.~4 hints at a power law 
$\langle m\rangle|_{p=p_c} \sim N^{-\eta_0}$ with $\eta_0 = 0.0356(8) > \eta_+$ (see also SM). 
The same is qualitatively true for the other models, where always $\eta_0 > \eta_+$ (see Table~1).
We see therefore that $z=0$ is no longer in the realm of uniform pointwise convergence to the FSS ansatz, 
and that therefore $z_0=0$. We should finally mention that we used $\Theta=0.5$ in Fig.~4, a value 
in between $\Theta_1$ and $\Theta_2$, as it gives the most systematic behavior for $z<0$. 
The same is true the other off-lattice models (but not for the 2d model, see SM), whence we conjecture that 
$\Theta=0.5$ for them.

The singularity of $g(z)$ at $z=0$ implies also that one cannot expect the effective 
critical points to scale as $p_c(N)-p \sim N^{-\Theta}$. Results obtained for the da Costa model, 
with $p_c(N)$ defined via equal peak masses, are shown in the SM. They indicate that 
$p_c(N)-p \sim N^{-\delta}$ with $\delta =0.9(1) > \Theta$. The agreement with the prediction 
$\delta=0.818(1)$ of \cite{Costa-2010} -- based on ``standard scaling relations" -- seems fortitious.

In this paper we do not present a detailed theory for the convergence to $g(z)$
for $z\leq 0$, in particular we do not explain how $\eta_0$ and $\delta$ are 
related to the exponent $\eta_-$. It could be that such a theory can be formulated more 
easily using either $\langle \log s_{\rm max} \rangle$ or 
$\langle 1/s_{\rm max}\rangle$ as an order parameter. But this would be beyond the scope of the present paper.

\begin{figure}
\begin{center}
\psfig{file=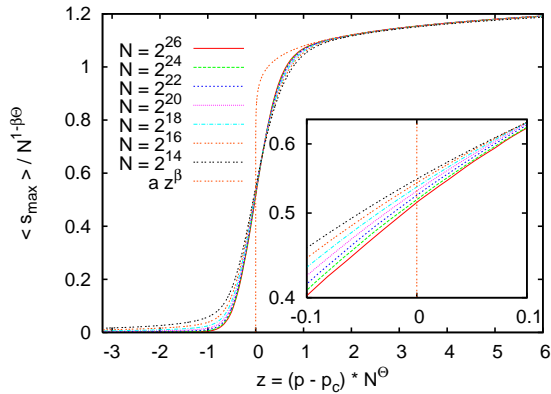,width=5.3cm, angle=270}
\caption{(Color online) Doubly linear plot of the same data shown in Fig.~3, but extended to values $p<p_c$.
Here we used $\Theta=1/2$, which gives worse data collapse for $p>p_c$, but vastly more systematic
behavior for $p<p_c$. The inset shows a blow-up of the region around $p=p_c$, with logarithmic y-axis.
The decrease of the curves at $z=0$ with $N$ suggests that $z_0=0$, and that a new power law holds for $p=p_c$.}
\end{center}
\end{figure}

In summary, we have shown that at least four models of explosive percolation, including the original 
product rule of Achlioptas {\it et al.} \cite{Achli-2009}, have continuous transitions. Each is in a different 
universality class, but all of them show unusual finite size behavior with a 
non-analytic scaling function. They all show double-peaked order parameter distributions with the sharpness
of the peaks increasing with system size, and different scaling laws for the width of the scaling region (
$\sim N^{-\Theta}$) and for the shift of the effective $p_c(N)$.  It would be 
interesting to see whether similar scaling holds in other percolation models with supposedly discontinuous 
transitions that are not explicitly related to Achlioptas-like dynamics 
\cite{Manna-2009,Araujo-2010,Buldy-2010}. It could be that the features 
found in the present paper arise from the specific non-locality of the Achlioptas process, and that this 
is why it was not seen previously in other critical phenomena.

We are indebted to Bob Ziff and Liang Tian for most useful correspondence.


\bibliography{PRL_bibliography}

\end{document}


\title{Supplementary Material for ``Explosive Percolation is Continuous, but with Unusual Finite Size Behavior"}
\author{Peter Grassberger} \affiliation{FZ J\"ulich, D-52425 J\"ulich, Germany} \affiliation{Complexity Science Group, University of Calgary, Calgary T2N 1N4, Canada}
\author{Claire Christensen} \affiliation{Complexity Science Group, University of Calgary, Calgary T2N 1N4, Canada}
\author{Golnoosh Bizhani} \affiliation{Complexity Science Group, University of Calgary, Calgary T2N 1N4, Canada}
\author{Seung-Woo Son} \affiliation{Complexity Science Group, University of Calgary, Calgary T2N 1N4, Canada}
\author{Maya Paczuski} \affiliation{Complexity Science Group, University of Calgary, Calgary T2N 1N4, Canada}
\date{\today}

\maketitle

\section{Simulation Details}

All simulations reported in the paper were made using modified versions of the fast Newman-Ziff algorithm
\cite{Newman-2001}, on a Linux workstation cluster. System sizes varied between $N = 2^{10}$ and $N =2^{26} 
(\approx 6.7\times 10^7)$ ($N$ is the number of nodes). For the smaller systems $\approx 10^8$ realizations 
were made for each model,
and for the largest systems this number was still $>10^4$. The control parameter $p$ is defined as in 
the references where the 4 models were introduced, as $p = L/N$ where $L$ is the number of links. 

Data were actually collected for fixed $n$, where $n$ is the number of clusters -- more precisely, in 
order to reduce the data files, $n$ was binned (typically with $\Delta n =1$ for smallest $N$ and 
$\Delta n =2^8$ for largest $N$). The values of $p$ quoted in the paper are average values over these
bins. Since most clusters in all four models are trees, except when $p$ is very large, there are 
very small fluctuations of $p$ for fixed $n$, and $\langle p\rangle$ depends smoothly on $n$. 
Moreover, test runs showed that the dependence of $s_{\rm max}$ on $n$ is at least as crisp as the 
dependence on $p$, i.e. $n$ is actually the more relevant control parameter. 

Mass distributions (Figs. 1 and 2 in the main paper) are obtained by binning, with typically 200 to 500
bins, and with bin sizes slowly increasing with $s_{\rm max}$ in order to take into account that the 
left hand peaks in Fig. 1 are sharper than the right hand peaks. For small $N$ the distributions shown 
in Figs. 1 and 2 are the raw data, modified just by interpolating between neighboring $n$ bins to 
obtain exactly equally high peaks (usually, no bin will have two peaks which have exactly equal height;
by ``interpolating" we mean taking weighted linear averages of the two histograms) 
and by normalizing them. For the largest $N$ this would have given 
too noisy plots, and cubic splines were used to make the plots more smooth.  

Unless otherwise noted, $p_c$ values are those in Table 1. They were determined by having best 
power laws $\sim N^{-\eta_0}$ for $\langle m \rangle$ at $p=p_c$.
 
\section{Modified Finite Size Scaling}

The finite size scaling ansaetze Eqs. (1) and (2) are of course never exact, and are usually understood
as 
\be
   \lim_{N\to\infty} N^{-\eta} P_{p=p_c,N}(m=z/N^\eta) = f(z)
\ee
and 
\be
   \lim_{N\to\infty} (p-p_c = z/N^\Theta)^{-\beta} \langle m\rangle = g(z)
\ee
for any fixed finite value of $z$. The limits here are pointwise limits, i.e. the norms of the 
differences between left and right hands converge uniformly to zero in any finite interval of $z$.
Furthermore, $f(z)$ and $g(z)$ are usually analytic (holomorphic) for all finite $z$. 

In a typical first order (discontinuous) transition, an attempt to construct $f(z)$ would give 
a function with two $\delta$-peaks. In that case the convergence could at best be {\it weak}, 
i.e. in distribution sense. Usually one prefers to call this not finite size scaling at all, 
although this is strictly spoken a matter of taste and convention.

In explosive percolation one has a ``mixed" situation: For the right hand peaks in Figs. 1 and 2
of the main paper one has pointwise convergence, if one chooses $\eta=\eta_+$. But then the 
left hand peak converges to a $\delta$-peak, i.e. the entire function $f(z)$ is approached 
only in the weak sense. Similarly, for $g(z)$ the convergence is strong (and $g(z)$ is 
analytic) for $z>0$ (strict inequality!), where only the right hand peak of $f(z)$ contributes.
The function $g(z)$ must vanish identically when only the left ($\delta$-) peak contributes,
which means that it must have a singularity at $z_0 \leq 0$, and convergence can only be weak 
in any interval containing $z_0$. As for first order transitions, it is a matter of convention
whether one calls this finite size scaling at all (as we did in this paper). 

\section{Analoga to Fig. 3 (main paper) for the other three models}

In the main paper, we showed in Fig. 3 for the da Costa model how $\langle m\rangle$ scales for 
$p>p_c$, and we said that a data collapse similar to that shown in the inset holds also for 
the other two off-lattice models, while the collapse is much worse for the 2d model. We now
show these data in Figs. S1 to S3.

\begin{figure}[htbp]
\begin{center}
\psfig{file=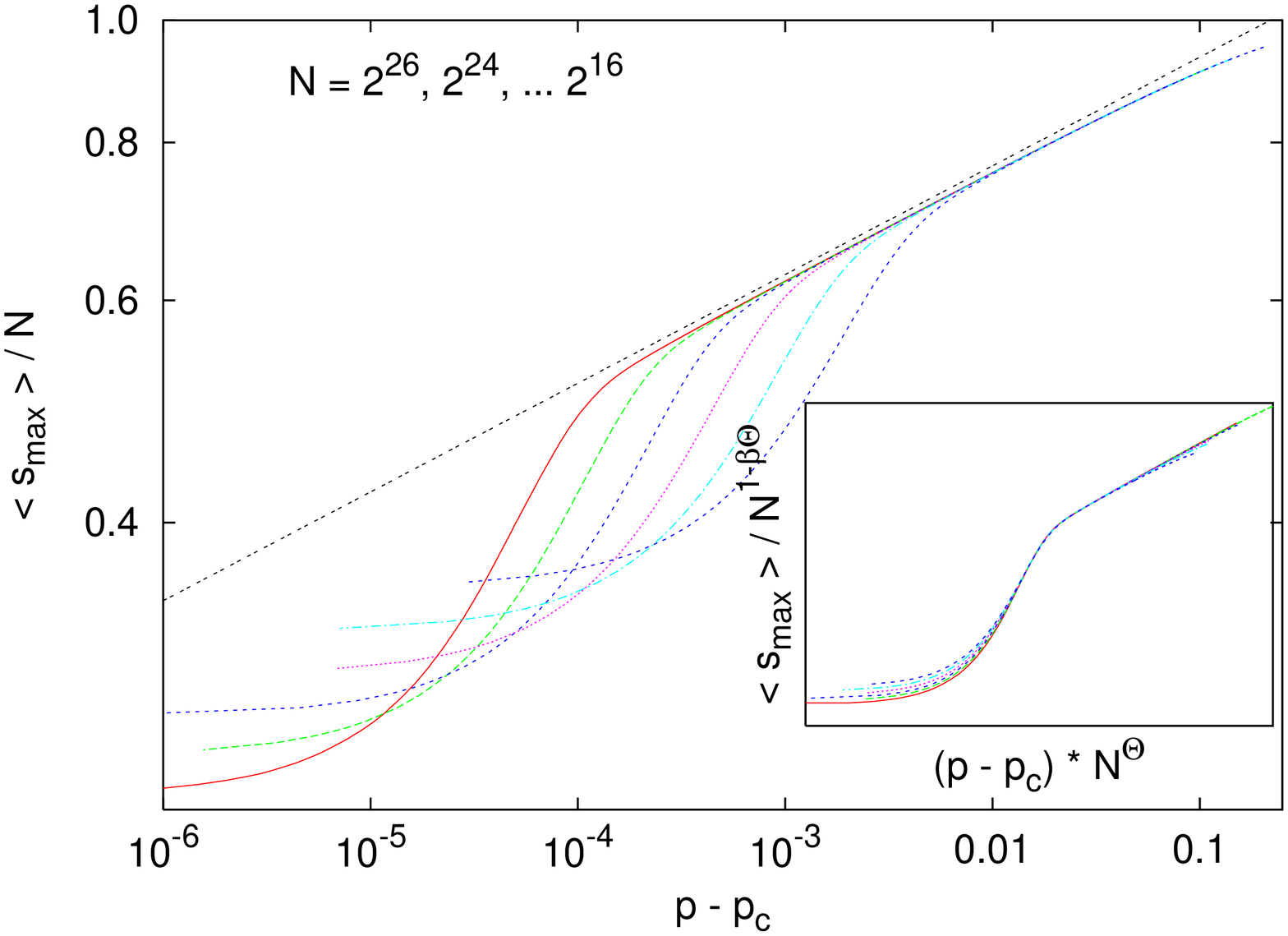, width=8.3cm, angle=270}
\caption{(Color online) Log-log plot of the average order parameter for the PR model {\it versus} $p-p_c$,
for six different values of $N$, similar to Fig. 3 of the main paper. The value of $p_c$ is chosen 
such that decrease of $\langle m\rangle$ with $N$, for $p\to p_c$, is a pure power law.}
\end{center}
\end{figure}

\begin{figure}[htbp]
\begin{center}
\psfig{file=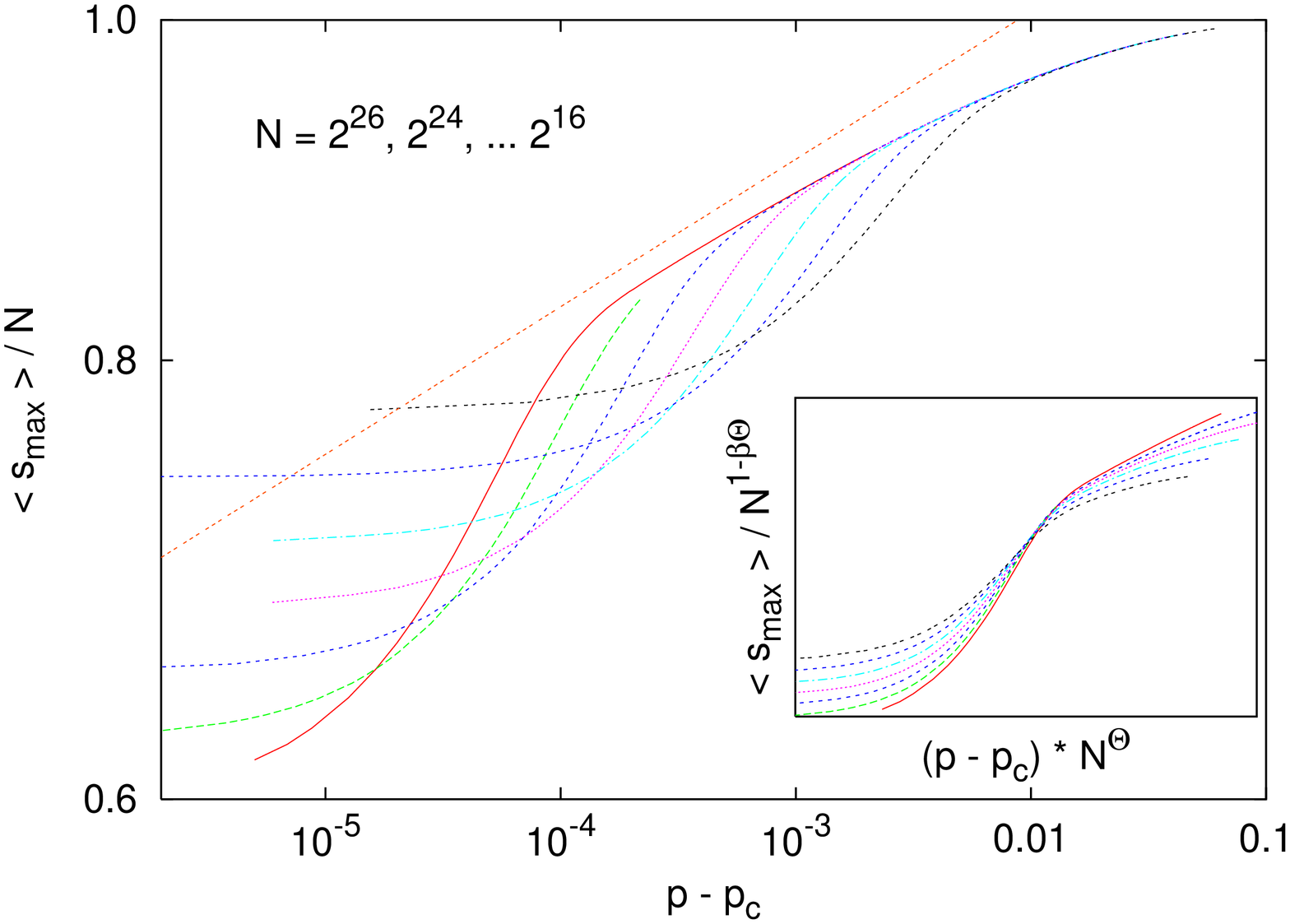, width=8.3cm, angle=270}
\caption{(Color online) Same as Fig. S1, but for the 2d model. This time the data collapse is much
worse (see the inset). For large $z\equiv(p-p_c)N^\Theta$ this is due to the much slower drift
of the right hand peak in Fig. 1. For $z\to 0$ it reflects the large difference between $\eta_0$
and $\eta_+$.}
\end{center}
\end{figure}

\begin{figure}[htbp]
\begin{center}
\psfig{file=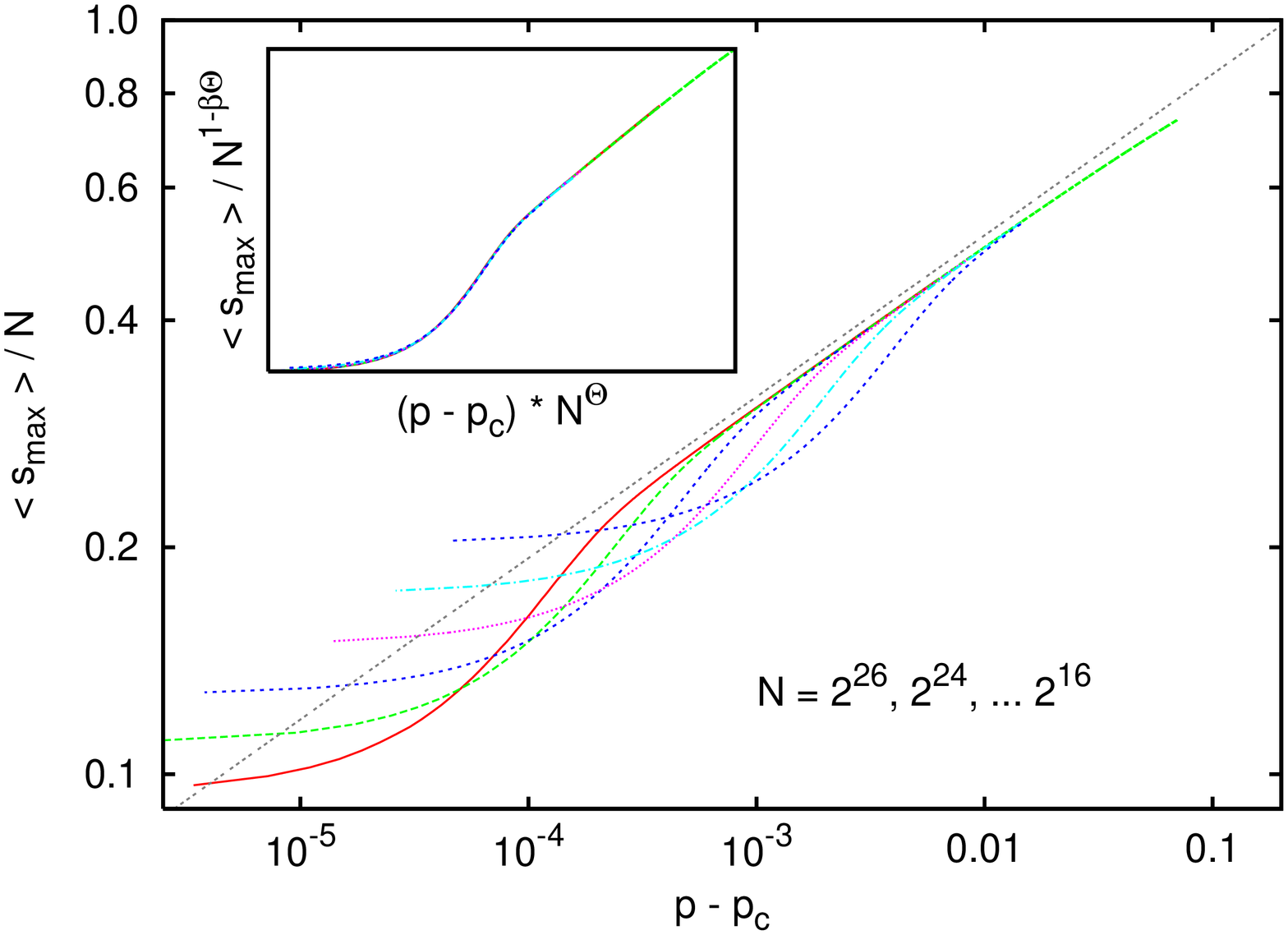, width=8.3cm, angle=270}
\caption{(Color online) Same as Fig. S1, but for the AE model. This time the data collapse is better
than in the other models (see the inset), reflecting the fact that the AE model is closest to 
an ordinary second order transition, among the four models studied here.}
\end{center}
\end{figure}

\section{Various other plots for the da Costa model}

Although the scaling behavior of the order parameter with $N$ at $p=p_c$ can, in principle, be 
inferred from Fig. 4 (main paper), we show the data also explicitly in Fig. S4. In this figure
we use three possible values of $p_c$ (one of them being the value obtained in \cite{Costa-2010},
in order to show how strongly the exponent $\eta_0$ depends on $p_c$.

\begin{figure}[htbp]
\begin{center}
\psfig{file=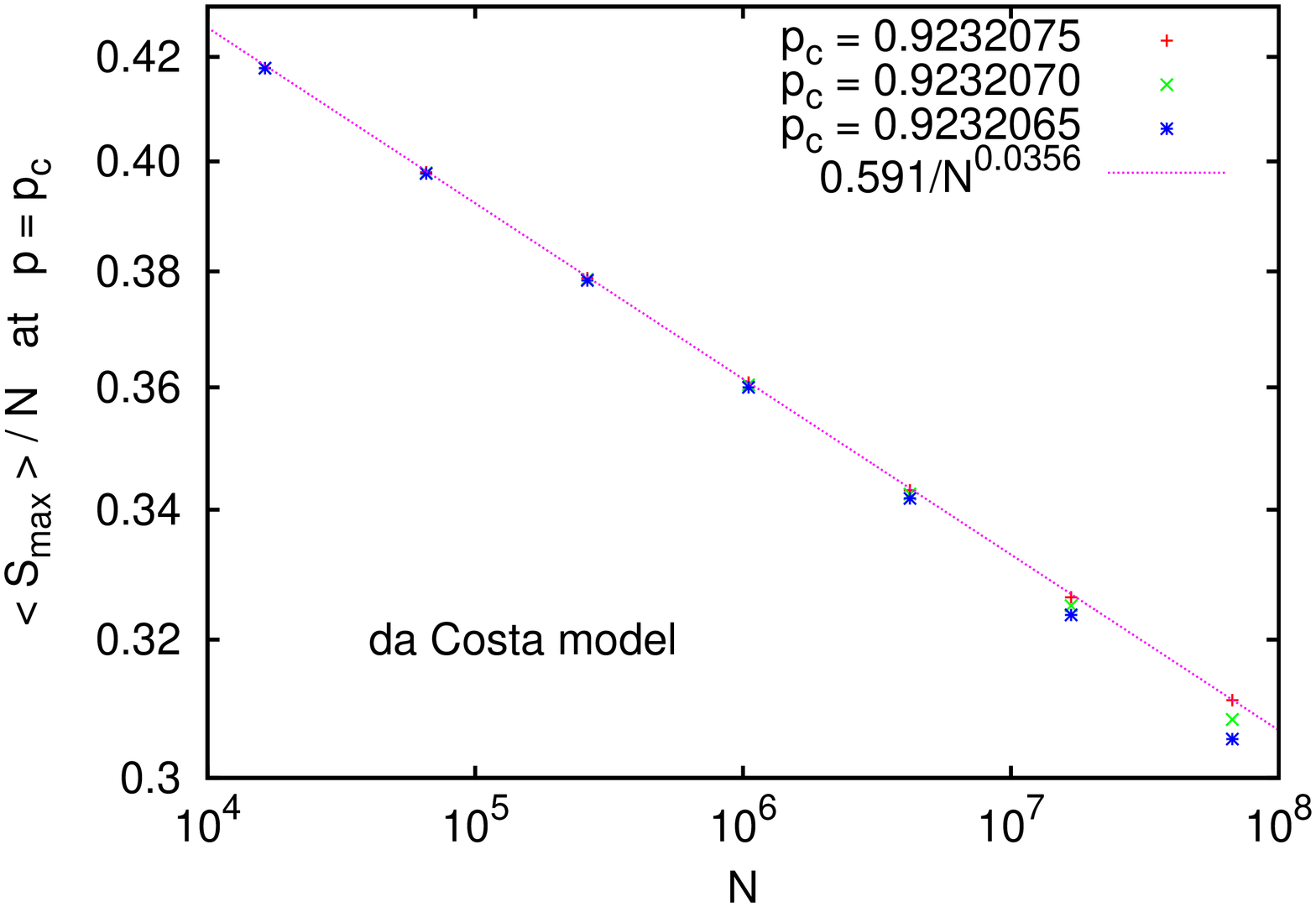, width=8.3cm, angle=270}
\caption{(Color online) Log-log plot of $\langle m\rangle$ at $p=p_c$ plotted {\it versus} $N$. In
order to show the sensitivity of the exponent $\eta_0$ to the precise value of $p_c$, curves for 
three values of the latter are shown.}
\end{center}
\end{figure}

Results for $p_c(N)$, the effective critical points on finite systems, are given in Fig. S5. 
Notice that values of $p_c(N)$ depend crucially on the operational procedure used to define 
effective critical points. One possibility would be e.g. the point where the two peaks in 
$P(s_{\rm max})$ have equal height (Fig. 1). For the da Costa model, this would give non-monotonic 
dependence on $N$. More natural seems the definition via equal areas under the two peaks. Notice 
that this would give ambiguous results for the 2d and AE models, as there the dips between the peaks
are not very deep. But for the da Costa model this is unproblematic. Figure 5 shows that our best
estimate of $p_c$ is slightly below the value of \cite{Costa-2010}, giving thereby the largest 
contribution to the uncertainty of the exponent $\delta$ (this slight inconsistency is also the 
reason why we used also these smaller $p_c$ values in Fig. S4). 
 
\begin{figure}[htbp]
\begin{center}
\psfig{file=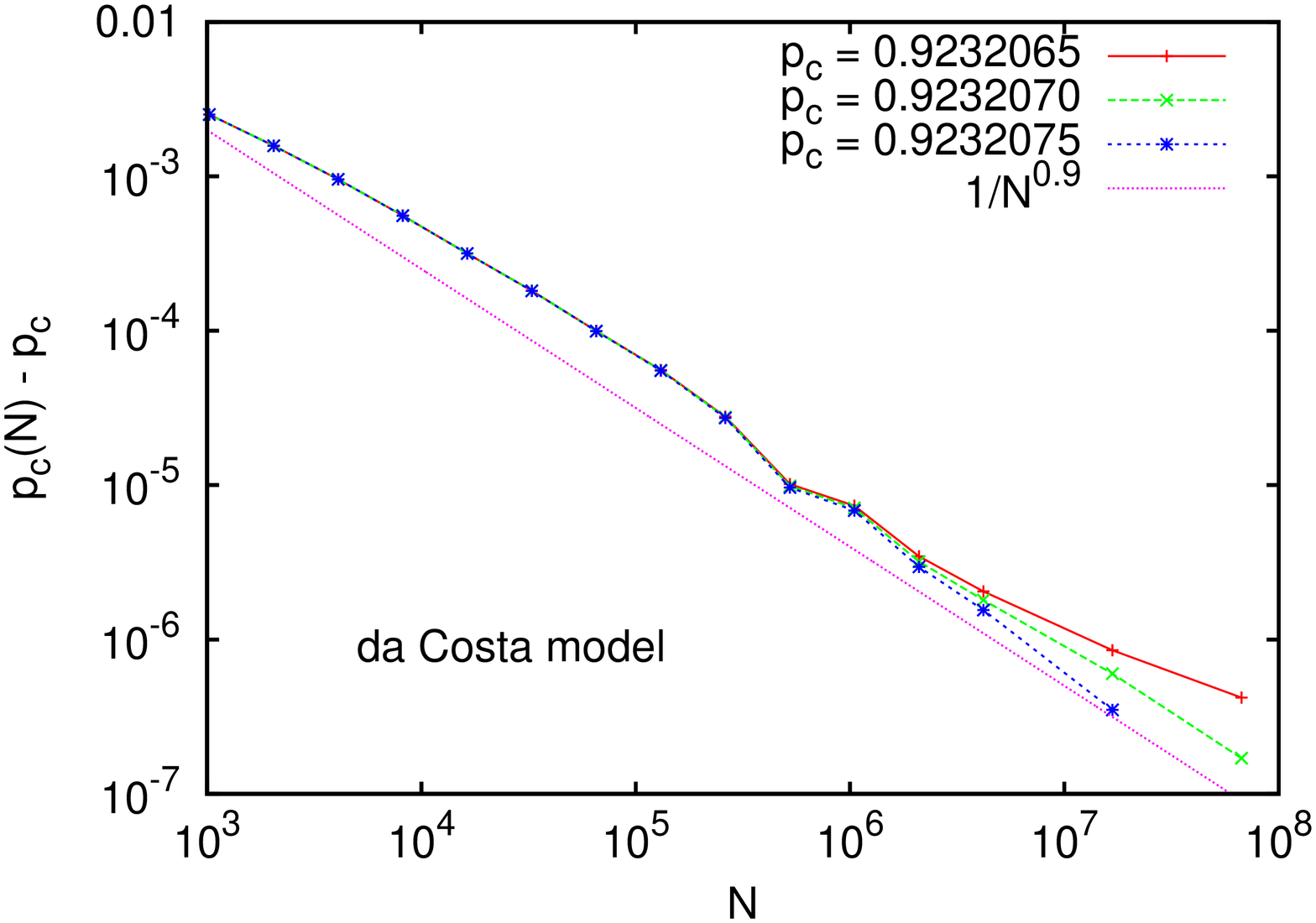, width=8.3cm, angle=270}
\caption{(Color online) Log-log plot of $p_c(N)$, defined as the value where the areas under 
both peaks in $P(s_{\rm max})$ have equal height, {\it versus} $N$. Notice that there are only 
two points for $N=2^{26}$, since our $p_c(N=2^{26})$ is smaller than the $p_c$ value of \cite{Costa-2010}.
}
\end{center}
\end{figure}

Alternatively, we could define $p_c(N)$ as the point where $P(s_{\rm max})$ has maximal variance.
Variances of $P(s_{\rm max})/N^{1-\eta_+}$ are plotted in Fig. S6 against $(p-p_c)N^{1/2}$. We see 
distributions which are for small $N$ markedly skewed and shifted away from the origin, but which 
become increasingly symmetric and centered at the origin as $N$ increases. Although this gives 
a much less precise estimate of $\delta$ than Fig. S5, it demonstrates also that $\delta > \Theta$.

\begin{figure}[htbp]
\begin{center}
\psfig{file=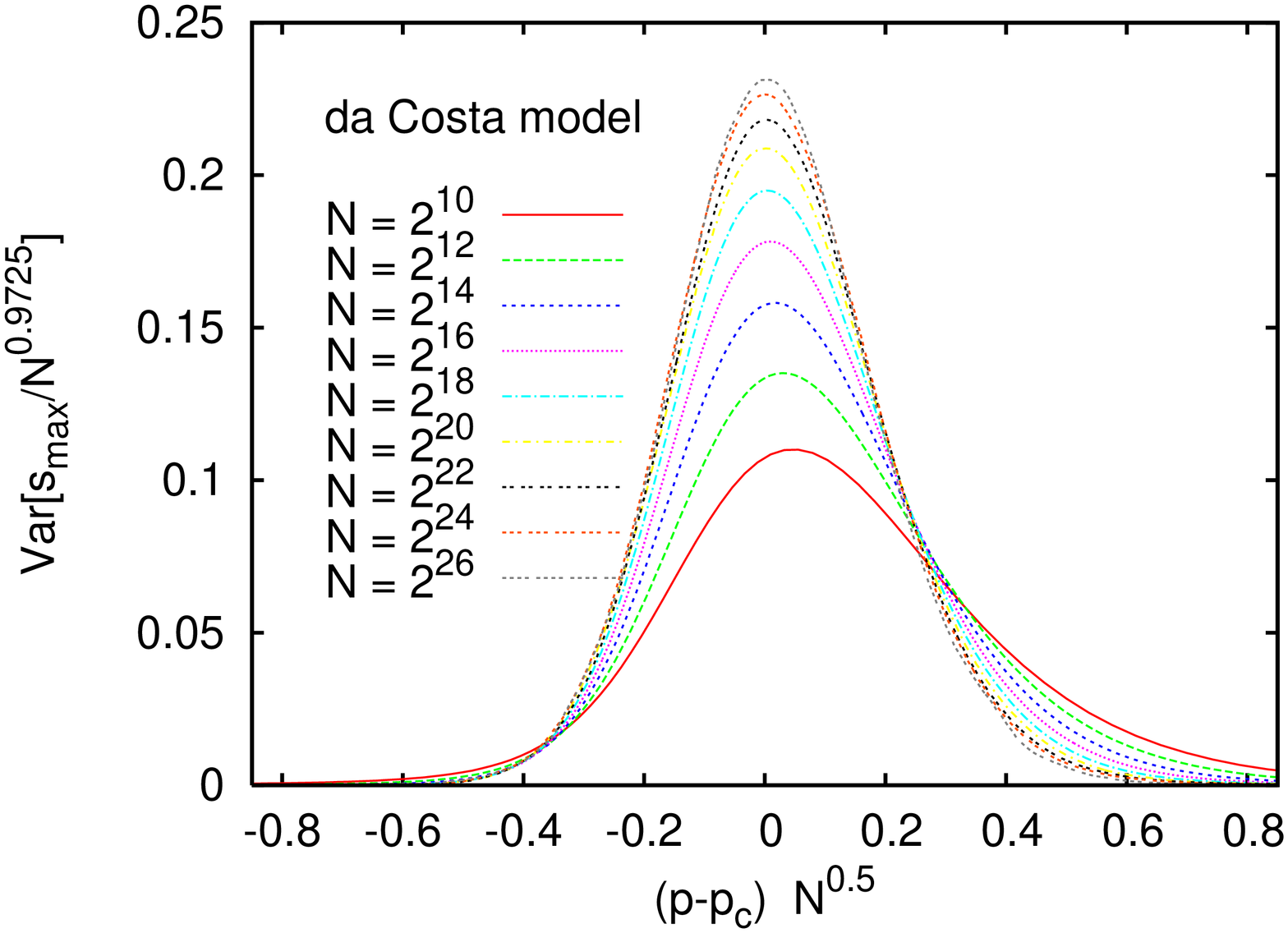, width=8.3cm, angle=270}
\caption{(Color online) Variances of $s_{\rm max}$ for fixed $N$, plotted {\it versus} $z = (p-p_c)N^{1/2}$,
using the $p_c$ value of \cite{Costa-2010}.
The width rescaling and the normalization are such that the curves should collapse for $N\to\infty$. 
For the $N$ values shown in the figure, the collapse is far from perfect, but this is to be expected. 
Notice that the horizontal peak positions hardly change for $N>2^{20}$, showing that $\delta > \Theta$.}
\end{center}
\end{figure}

Plots similar to Figs. S5 and S6 were not made for the other models, the main 
reason being the larger uncertainties of $p_c$.

\section{For the 2d model, $\Theta$ is strictly smaller than 1/2}

As we said in the paper, one observation that corroborates $\Theta=1/2$ for the off-lattice models
is that it gives very ``regular" behavior of $\langle m\rangle$ in the near subcritical region.
Instead of showing here the evidence for this, we show for the 2d model what can go wrong, if 
$\Theta$ is chosen badly. More precisely, we show in the top panel of Fig. S7 results for $\Theta=0.47$,
and results for $\Theta=0.5$ in the lower panel. It seems clear that panel (b) is not very plausible,
in particular since the scaling $m_- \sim N^{-\eta_-}$ with $\eta_-> \eta_+$ of the left hand peaks in 
Fig. 1 requires that the curves decrease with $N$ for $z<0$. For the other models similar curve 
crossings appeared when $\Theta = \Theta_2$ was used instead of $\Theta=1/2$.

\begin{figure}[htbp]
\begin{center}
\psfig{file=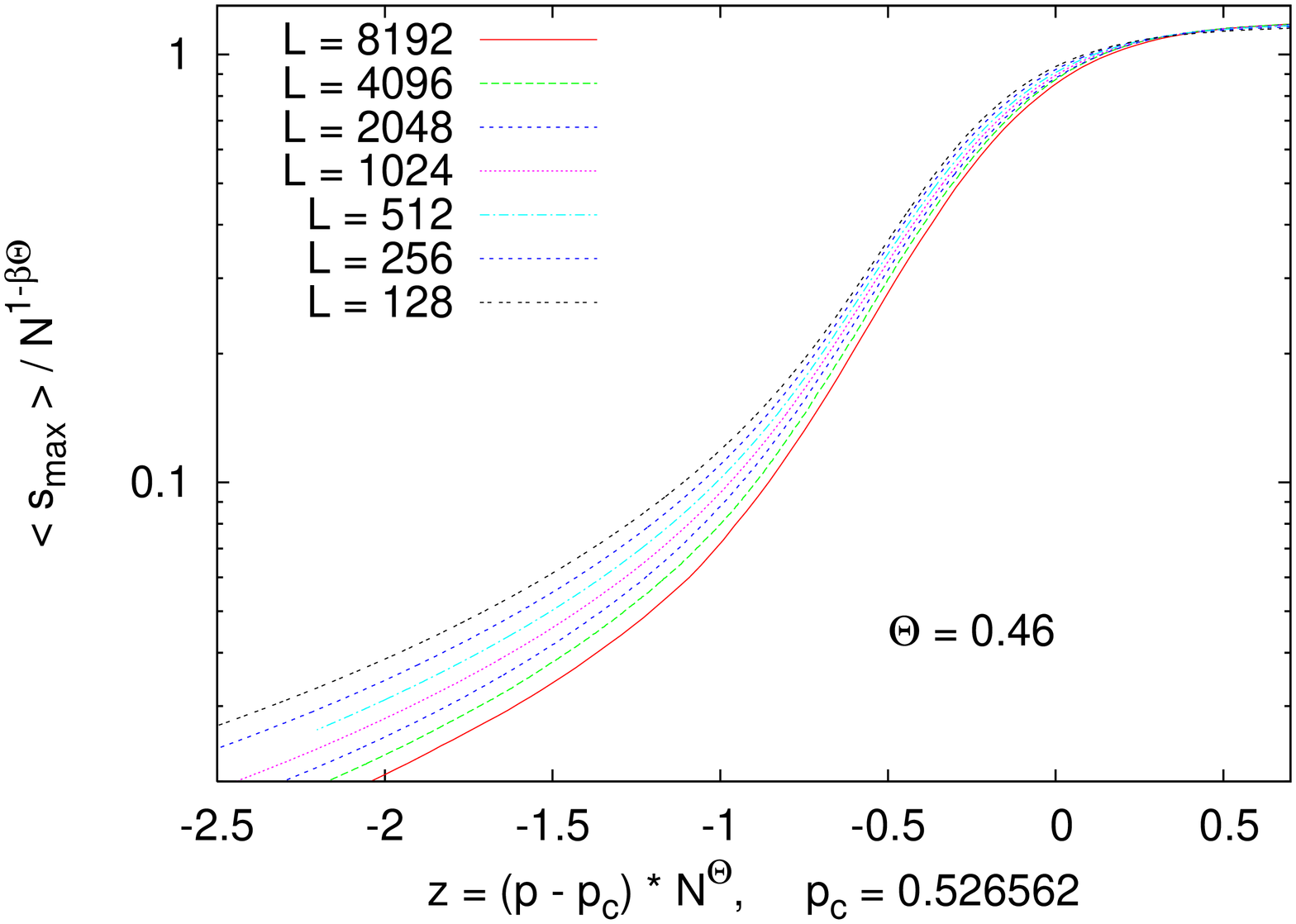, width=7.3cm, angle=270}
\psfig{file=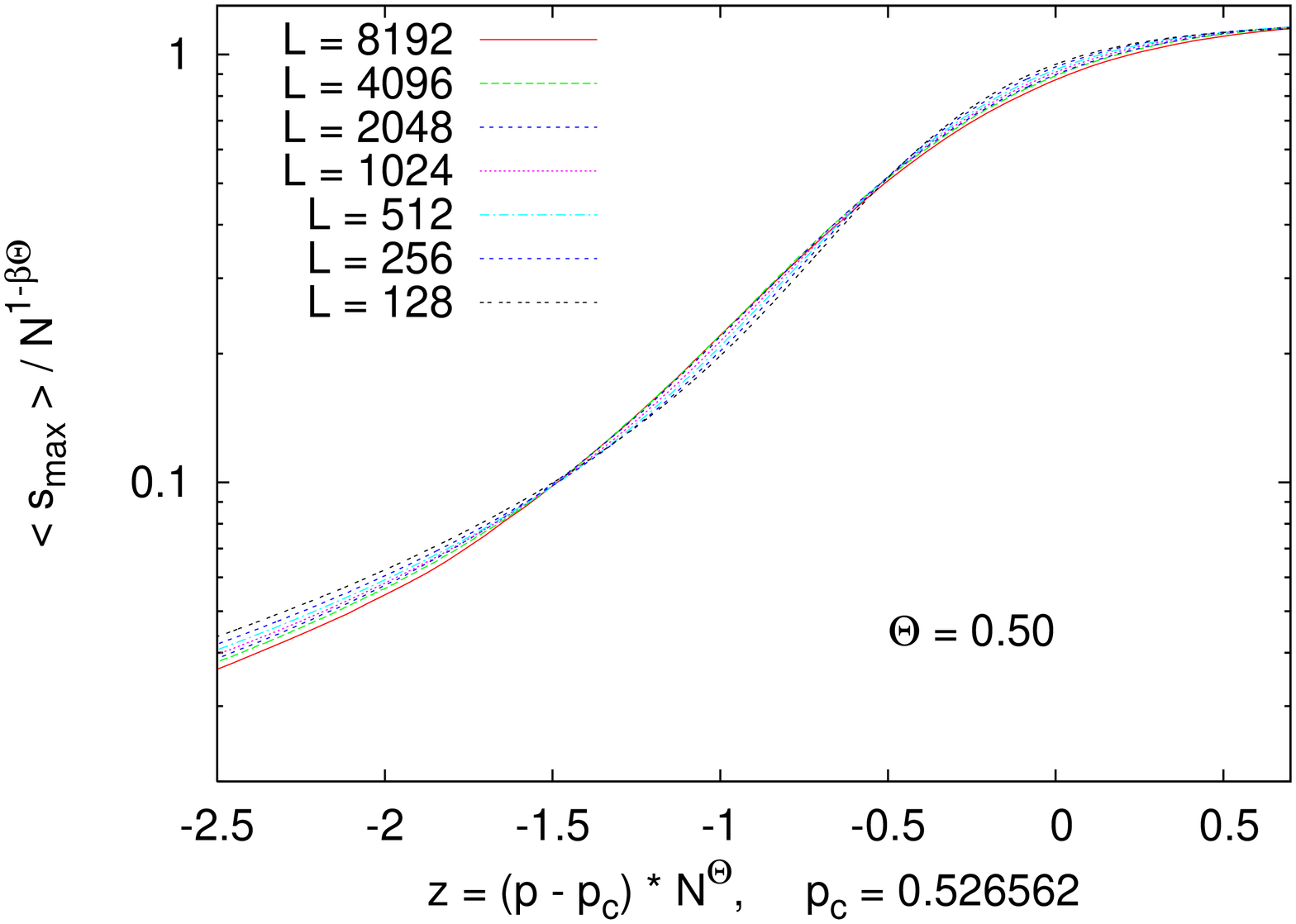, width=7.3cm, angle=270}
\caption{(Color online) Linear-log plots of $\langle s_{\rm max}\rangle / N^{1-\beta\Theta}$ against
  $(p-p_c)N^{\Theta}$ for the 2d model. In both plots we used the values for $\beta$ and $p_c$ given 
in Table 1. In the upper panel we used $\Theta = 0.46$ (the average between $\Theta_1$ and $\Theta_2$),
while we used $\Theta=0.5$ in the lower panel.}
\end{center}
\end{figure}
  
\section{Further comparisons with previous papers}

a) To our knowledge, the only previous work where $\Theta \neq \delta$ was seen in a continuous phase 
transition is \cite{Prellberg}. This dealt with interacting self avoiding walks in 4 dimensions.
The authors also found double-peaked distributions of the order parameter, but they did not 
report different scaling laws for both peaks. Thus the reason for $\Theta \neq \delta$ was not 
explained, as in our case, via a singularity of the scaling function $g(z)$. Also, in 
\cite{Prellberg} the authors found $\Theta \gg \delta$, while we found in our models $\Theta < \delta$.

b) A different scaling theory for the PR model was presented in \cite{Cho-2010}. The authors
there started from the assumption that $\langle m\rangle$ is independent of $N$ at $p=p_c$, which 
is definitely not true for any of the four models according to our simulations. Although this 
prevents our theories from being equivalent, there are some similarities. In particular, 
the authors of \cite{Cho-2010} show that the width of the FSS region scales as $N^{-\theta}$ with 
$\theta=0.48$, which is very close to our conjectured value $\Theta=1/2$. 

c) A detailed study of the 2d model was made in \cite{Ziff-2010}. The most remarkable agreement
is that $\eta_0$ was measured there as $0.0589(10)$, while we found $0.0612(8)$. The slight 
discrepancy is partially due to a slightly different estimate of $p_c$ ($0.526565(5)$ in 
\cite{Ziff-2010} against $0.526562(3)$ in the present work). Also the estimate of the 
Fisher exponent $\tau$ in \cite{Ziff-2010} is fully compatible with our value of $\beta$, if 
we accept the relationship between the two exponents given in \cite{Costa-2010}.

\bibliography{PRL_bibliography}